\documentclass[floats, prd, eqnum, showpacs, nofootinbib, twocolumn,eqsecnum]{revtex4-1}

\usepackage{color,graphicx}
\usepackage{amsfonts}
\usepackage{amssymb}
\usepackage{comment}
\begin{document}

\title{Linearization stability of reflection-asymmetric thin-shell wormholes with double shadows}
\author{Naoki Tsukamoto${}^{1}$}\email{tsukamoto@rikkyo.ac.jp}

\affiliation{
${}^{1}$Department of General Science and Education, National Institute of Technology, Hachinohe College, Aomori 039-1192, Japan \\
}

\begin{abstract}
Wormholes are hypothetical objects which can be black hole mimickers with strong gravitational fields. 
Recently, Wielgus \textit{et al.} have 
constructed reflection-asymmetric thin-shell wormholes 
which are composed of the parts of a Schwarzschild spacetime and a Reissner-Nordstrom spacetime with two photon spheres with different sizes each other
in~[M.~Wielgus, J.~Horak, F.~Vincent, and M.~Abramowicz, Phys. Rev. D \textbf{102}, 084044 (2020)]. 
They have discussed observational property of the shadows with two photon rings in different sizes as seen from an observer and they have named their shadows double shadows. 
In this paper, we study the linearization stability of the reflection-asymmetric thin-shell wormholes with the double shadows.
\end{abstract}

\maketitle

\section{Introduction}
Recently, LIGO and VIRGO Collaborations have detected gravitational waves from black hole binaries~\cite{Abbott:2016blz} 
and Event Horizon Telescope Collaboration has reported the shadow image of a black hole candidate at the center of a giant elliptical galaxy M87~\cite{Akiyama:2019cqa}.
The theoretical and observational aspects of compact objects in general relativity will be more important than before.  

Wormholes are hypothetical objects with a non-trivial topology in general relativity~\cite{Visser_1995,Morris:1988cz}.
Morris and Thorne have discussed the passability of wormholes 
and they have also shown that energy conditions violate at least at the throat of static and spherically symmetric wormholes if we assume general relativity without a cosmological constant~\cite{Morris:1988cz}.
The wormholes can be black hole mimickers because they can have strong gravitational fields. 
For example, spherically symmetric wormholes with strong gravitational fields 
can have unstable (stable) circular light orbits named photon spheres (antiphoton spheres)~\cite{Perlick:2003vg,Nandi:2006ds,Tsukamoto:2012xs,Tsukamoto:2016qro,Nandi:2018mzm,Shaikh:2018oul,Shaikh:2019jfr,Tsukamoto:2020uay,Tsukamoto:2020bjm}.

A thin-shell wormhole whose energy conditions are broken only at a throat which is supported by a thin shell~\cite{Lanczos:1922,Lanczos:1924,Israel:1966rt,Poisson:2004} 
was considered by Visser~\cite{Visser:1989kg} with Darmois-Israel matching~\cite{Darmois:1927,Israel:1966rt,Poisson:2004}.
The linearization stability of the thin shell of a Schwarzschild wormhole was investigated by Poisson and Visser~\cite{Poisson:1995sv} 
and then the stability of thin-shell wormholes such as 
Reissner-Nordstrom wormholes~\cite{Eiroa:2003wp,Eiroa:2007qz,Eiroa:2008ky,Eiroa:2009hm,
Kuhfittig:2010pb,Sharif:2013lna,Sharif:2013tva,Eid:2016axb,HabibMazharimousavi:2017qfb,Wang:2017gdp,Forghani:2019wgt},
the other static and spherically symmetric wormholes~\cite{Kim:1992sh,Barcelo:2000ta,Ishak:2001az,Lobo:2003xd,Lobo:2005zu,Eiroa:2005pc,Lobo:2005yv,Rahaman:2006xb,Rahaman:2007bf,Eiroa:2007qz,Eiroa:2008ky,Eiroa:2009hm,Garcia:2011aa,Eiroa:2015hrt},
plane symmetric wormholes~\cite{Lemos:2008aj},
cylindrical symmetric wormholes~\cite{Eiroa:2004at,Bejarano:2006uj},
higher-dimensional wormholes~\cite{Thibeault:2005ha,Rahaman:2006vg,Mazharimousavi:2010bm,Kokubu:2014vwa,Mehdizadeh:2015dta,Kokubu:2015spa}, 
lower-dimensional wormholes~\cite{Perry:1991qq,Kim:1993bz,Rahaman:2011yh,Eiroa:2013xj,Tsukamoto:2018lsg}, 
and wormholes in an expanding spacetime~\cite{LaCamera:2011qi} have been discussed.
The passability~\cite{Nakao:2013hba} and nonlinear stability~\cite{Akai:2017aro} of thin-shell wormholes have been studied. 

Usually two copied manifolds are used to construct the thin-shell wormholes. 
It would be difficult to distinguish a reflection-symmetric wormhole spacetime constructed by the parts of the copied black hole spacetimes 
from the original black hole spacetime by astronomical observations
if the wormhole has photon spheres and if there are few light sources in the other side of the throat.~\footnote{In principle, 
we would distinguish wormholes from black holes.
For an example, if there are stars bounded by a wormhole in both sides of the wormhole, 
the orbit of the star in one side can be affected by the gravity of the star of the other side.~\cite{Simonetti:2020vhw,Dai:2019mse}}

There were few researches on reflection-asymmetric thin-shell wormholes such as~\cite{Garcia:2011aa,Eid:2017xcg,Forghani:2018gza}.
Recently, the shadow of a reflection-asymmetric thin-shell wormhole which is composed of the parts of two Schwarzschild spacetime with different masses 
were discussed by Wang \textit{et al.}~\cite{Wang:2020emr}.
Wielgus \textit{et al.} have constructed a reflection-asymmetric thin-shell wormhole which is composed of 
the parts of the Schwarzschild spacetime with a photon sphere and a Reissner-Nordstrom spacetime with another photon sphere
and they have discussed the observational property of the shadow~\cite{Wielgus:2020uqz}.
They have concluded that the shadow of the asymmetric wormhole has two photon rings in different sizes as seen from an observer 
and they have named the shadow double shadow. 
The asymmetric thin-shell wormhole with the photon spheres
can be distinguished from the black holes by the observations since light rays can be reflected by a potential wall near the throat~\cite{Wielgus:2020uqz}.

In this paper, 
we investigate the linearization stability of the reflection-asymmetric thin-shell wormhole which is composed of 
the parts of the Schwarzschild spacetime and the Reissner-Nordstrom spacetime with the double shadow constructed in~\cite{Wielgus:2020uqz}.
 
This paper is organized as follows. 
We review the Reissner-Nordstrom spacetime very briefly in Sec.~II
and we construct the reflection-asymmetric thin-shell wormhole with the double shadow in Sec.~III.
We investigate the linearization stability of the reflection-asymmetric thin-shell wormhole in Sec.~IV 
and we conclude our results in Sec.~V.
In this paper we use the units in which a light speed and Newton's constant are unity.

\section{Reissner-Nordstrom spacetime}
A Reissner-Nordstrom spacetime has a line element
\begin{eqnarray}
ds^2 
=-f(r) dt^2 + \frac{dr^2}{f(r)}+r^2(d\theta^2 + \sin^2 \theta d\phi^2),
\end{eqnarray}
where $f(r)$ is given by
\begin{eqnarray}
f(r) \equiv 1-\frac{2M}{r} +\frac{Q^2}{r^2}
\end{eqnarray}
and where $M>0$ and $Q$ are a mass and a charge, respectively. 
It is a black hole spacetime with an event horizon at $r=r_{\mathrm{EH}}\equiv M+\sqrt{M^2-Q^2}$
for $Q^2 \leq M^2$
while it has naked singularity for $M^2 < Q^2$.
There is a photon sphere at
\begin{eqnarray}
r=r_{\mathrm{PS}}\equiv \frac{3M+\sqrt{9M^2-8Q^2}}{2}
\end{eqnarray}
for $Q^2 \leq  9M^2/8$
and there is an antiphoton sphere at  
\begin{eqnarray}
r=r_{\mathrm{APS}}\equiv \frac{3M-\sqrt{9M^2-8Q^2}}{2}
\end{eqnarray}
for $M^2 < Q^2 \leq  9M^2/8$.


\section{Reflection-Asymmetric thin-shell wormhole}
In this section, we construct a wormhole spacetime without reflection symmetry or Z2 symmetry 
by the Darmois-Israel matching~\cite{Darmois:1927,Israel:1966rt,Poisson:2004}.
We make two manifolds $\mathcal{M}_\pm \equiv \left\{ r>a  \right\}$, where $a$ is a constant satisfying $a > r_{\mathrm{EH}\pm}$ 
by removing $\Omega_\pm \equiv \left\{ r \leq  a  \right\}$ from the Reissner-Nordstrom spacetimes.
The boundaries of the manifolds~$\mathcal{M}_\pm$ are timelike hypersurfaces $\Sigma_\pm \equiv \left\{ r = a  \right\}$ 
and we identify the hypersurfaces $\Sigma \equiv \Sigma_+ = \Sigma_- $.
As a result, we obtain a manifold $\mathcal{M}$ by gluing the manifolds $\mathcal{M}_\pm$ at a throat located at $\Sigma$.
The hypersurface~$\Sigma$ is filled with a Dirac distribution matter and it is called thin shell.
We permit $a=a(\tau)$, where $\tau$ is the proper time of the thin shell since we are interested in the stability of the thin shell~$\Sigma$. 

The line elements in the domains $\mathcal{M}_\pm$ are given by 
\begin{equation}
ds_\pm^2 
=-f_\pm(r) dt^2_\pm + \frac{dr^2}{f_\pm(r)} +r^2 (d\theta^2 + \sin^2 \theta d\phi^2),
\end{equation}
where $f_\pm(r)$ are given by
\begin{eqnarray}
f_\pm(r) \equiv 1-\frac{2M_\pm}{r} +\frac{Q_\pm^2}{r^2}.
\end{eqnarray}
Note the time coordinates $t_\pm$ are discontinuous on the hypersurface $\Sigma$ 
while coordinates $r$, $\theta$, and $\phi$ are continuous across the hypersurface $\Sigma$.

We assume that we can set coordinates $y^i=(\tau, \theta, \phi)$ on the both sides of $\Sigma$.
Let the thin shell be at $t_\pm=T_\pm(\tau)$ and $r=a(\tau)$.
We set the unit normal vectors of the thin shells as
\begin{eqnarray}\label{eq:normalvector}
n_{\mu \pm} dx^\mu_\pm= \pm \left( -\dot{a} dt_\pm +\dot{T}_\pm dr \right),
\end{eqnarray}
where the overdot is a differentiation with respect to $\tau$.
The four velocity of the thin shell is given by
$u^{\mu}_{\pm} \partial_{\mu \pm}=\dot{T}_\pm \partial_{t\pm}+\dot{a} \partial_r$. 
From the normalization of the four velocity $u^\mu_\pm u_{\mu \pm}=-1$, 
we obtain
\begin{equation}\label{eq:uu}
f_\pm (a) \dot{T}^2_\pm - \frac{\dot{a}^2}{f_\pm(a)}=1,
\end{equation}
where $\dot{T}_\pm$ should be
\begin{eqnarray}
\dot{T}_\pm= \frac{\sqrt{f_\pm+\dot{a}^2}}{f_\pm}.
\end{eqnarray}
By using Eq.~(\ref{eq:uu}) and the basis vectors $e^\mu_{i \pm} \equiv \partial x^{\mu}_{\pm} /\partial y^{i}$ given by
\begin{eqnarray}
&&e^\mu_{\tau \pm} \partial_{\mu \pm} =\dot{T}_\pm \partial_{t\pm} + \dot{a} \partial_r, \\
&&e^\mu_{\theta \pm} \partial_{\mu \pm} = \partial_\theta,  \\
&&e^\mu_{\phi \pm} \partial_{\mu \pm} = \partial_\phi,
\end{eqnarray}
the induced metric $h_{ij \pm}\equiv g_{\mu \nu \pm} e^{\mu}_{i \pm} e^{\nu}_{j \pm}$ 
on the hypersurface $\Sigma$ in $\mathcal{M}_\pm$ is given by
\begin{eqnarray}
ds_{\Sigma}^2
&=&\left. ds_\pm^2 \right|_{\Sigma}\nonumber\\
&=&h_{ij \pm}dy^idy^j \nonumber\\
&=&-d\tau^2+a^2 \left( d\theta^2 +\sin^2 \theta d\phi^2  \right).
\end{eqnarray}
It guarantees that the metric on the hypersurface $\Sigma$ is the same as viewed from both sides.

The thin shell satisfies Einstein equations 
\begin{equation}\label{eq:Einstein_eq}
S^i_j= - \frac{1}{8\pi} \left( \left[ K^i_j \right] - \left[ K \right] \delta^i_j \right),
\end{equation} 
where the bracket $\left[ \textrm{\boldmath $F$} \right]$ denotes the jump of any function $\textrm{\boldmath $F$}$ across $\Sigma$,
\begin{equation}
\left[ \textrm{\boldmath $F$} \right] \equiv \left. \textrm{\boldmath $F$}_+ \right|_{\Sigma} - \left. \textrm{\boldmath $F$}_- \right|_{\Sigma},
\end{equation}
where $\textrm{\boldmath $F$}_+$ and $\textrm{\boldmath $F$}_-$ are $\textrm{\boldmath $F$}$ in $\mathcal{M}_+$ and $\mathcal{M}_-$, respectively,
and $S^i_j$ is the surface stress-energy tensor of the thin shell given by
\begin{equation}\label{eq:S_ij}
S^i_j=(\sigma+p)U^iU_j+p \delta^i_j,
\end{equation}
where $U_i$ is given by $U_i dy^i \equiv u_{\mu \pm} e^\mu_{i \pm} dy^i= -d\tau$, 
and where $\sigma=-S^\tau_\tau$ and $p=S^\theta_\theta=S^\phi_\phi$ are the surface energy density and the surface pressure of the thin shell, respectively.
Here,  $K_{ij}$ is the extrinsic curvature given by
\begin{equation}\label{eq:K_ij}
K_{ij}
\equiv n_{\mu;\nu} e^\mu_i e^\nu_j,
\end{equation}
where $;$ is the covariant derivative.
By using the normal vectors~(\ref{eq:normalvector}), the extrinsic curvatures of the hypersurfaces in $\mathcal{M}_\pm$ are given by 
\begin{eqnarray}\label{eq:K_tautau}
K^\tau_{\tau \pm}&=& \frac{\pm 1}{\sqrt{\dot{a}^2+f_\pm}}\left( \ddot{a}+\frac{f^{\prime}_\pm}{2} \right), \\\label{eq:K_thetatheta}
K^\theta_{\theta \pm}&=&K^\phi_{\phi \pm}= \frac{\pm \sqrt{\dot{a}^2+f_\pm}}{a}, 
\end{eqnarray}
and the traces are obtained as
\begin{equation}\label{eq:K}
K_\pm = \frac{\pm 1}{\sqrt{\dot{a}^2+f_\pm}}\left( \ddot{a}+\frac{f^{\prime}_\pm}{2} \right) \pm \frac{2}{a}\sqrt{\dot{a}^2+f_\pm}.
\end{equation}
From $(\tau,\tau)$ and $(\theta,\theta)$ components of the Einstein equations~(\ref{eq:Einstein_eq}), we obtain
\begin{equation}\label{eq:S^tau_tau}
\sigma= - \frac{\sqrt{\dot{a}^2+f_+}}{4\pi a} -\frac{\sqrt{\dot{a}^2+f_-}}{4\pi a}
\end{equation}
and
\begin{eqnarray}\label{eq:S^phi_phi}
p&=&\frac{1}{8\pi \sqrt{\dot{a}^2+f_+}} \left( \ddot{a} + \frac{\dot{a}^2+f_+}{a} +\frac{f^{\prime}_+}{2} \right) \nonumber\\
&&+\frac{1}{8\pi \sqrt{\dot{a}^2+f_-}} \left( \ddot{a} + \frac{\dot{a}^2+f_-}{a} +\frac{f^{\prime}_-}{2} \right)
\end{eqnarray}
and then we get, from Eqs. (\ref{eq:S^tau_tau}) and (\ref{eq:S^phi_phi}),
\begin{equation}\label{eq:Energy_conservation0}
\frac{d( \sigma \mathcal{A})}{d\tau} +p\frac{d \mathcal{A}}{d\tau}=0,
\end{equation}
where $\mathcal{A} \equiv 4 \pi a^2$ is the area of the throat. 
Equation~(\ref{eq:Energy_conservation0}) can be expressed by
\begin{equation}\label{eq:Energy_conservation2}
a\sigma^{\prime}+2(\sigma +p) =0,
\end{equation}
where the prime denotes the differentiation with respect to $a$ and $\sigma^{\prime}=\dot{\sigma}/\dot{a}$. 
We assume that the thin shell is filled with a barotropic fluid with $p=p(\sigma)$.
From Eq.~(\ref{eq:Energy_conservation2}), we notice that the surface density of the barotropic fluid is expressed as a function of $a$ or $\sigma=\sigma(a)$.
The equation of motion of the thin shell is given by, from Eq.~(\ref{eq:S^tau_tau}),
\begin{equation}
\dot{a}^2+V(a)=0,
\end{equation}
where $V(a)$ is an effective potential defined by
\begin{equation}
V(a) \equiv  \bar{f} -\left( \frac{\Delta}{4\pi a \sigma} \right)^2 -\left( 2\pi a \sigma \right)^2,
\end{equation}
where $\bar{f}$ and $\Delta$ are given by
\begin{eqnarray}
\bar{f} \equiv \frac{f_{-}+f_{+}}{2}
\end{eqnarray}
and
\begin{eqnarray}
\Delta \equiv \frac{f_{+}-f_{-}}{2},
\end{eqnarray}
respectively.
The derivative of $V$ with respect to $a$ is obtained as
\begin{equation}
V^{\prime}= \bar{f}^{\prime} -\frac{\Delta\left[ \Delta^{\prime}a\sigma -\Delta(\sigma +a\sigma^{\prime}) \right]}{8\pi^2 a^3 \sigma^3} -8\pi^2 a \sigma (\sigma +a\sigma^{\prime})
\end{equation}
and, from Eq.~(\ref{eq:Energy_conservation2}), it can be rewritten as
\begin{equation}
V^{\prime}= \bar{f}^{\prime} -\frac{\Delta\left[ \Delta^{\prime}a\sigma +\Delta(\sigma +2p) \right]}{8\pi^2 a^3 \sigma^3} +8\pi^2 a \sigma (\sigma +2p).
\end{equation}
By using Eq.~(\ref{eq:Energy_conservation2}) again, the second derivative of $V$ is obtained as
\begin{eqnarray}
V'' =
&&\bar{f}^{\prime \prime} -\frac{\Delta^{\prime 2}}{8\pi^2 a^2 \sigma^2} 
-\frac{\Delta}{8\pi^2 a^4 \sigma^4} \left[ 4\Delta^{\prime}(\sigma +2p) a \sigma \right. \nonumber\\
&&\left.+ \Delta'' a^2 \sigma^2 -2\Delta\sigma(\sigma +p)(1+2\beta^2)+3\Delta(\sigma +2p)^2 \right] \nonumber\\
&& -8\pi^2 \left[ (\sigma +2p)^2 +2\sigma (\sigma +p)(1+2\beta^2) \right],
\end{eqnarray}
where $\beta^2\equiv dp/d\sigma=p^{\prime}/\sigma^{\prime}$.

Here and hereafter, we impose a constraint 
\begin{eqnarray}\label{eq:constraint}
f_+(a)=f_-(a)
\end{eqnarray}
and we concentrate on the case that the manifold $\mathcal{M}_-$ is the part of the Schwarzschild black hole spacetime, i.e., $Q_-=0$,
as well as Ref.~\cite{Wielgus:2020uqz}.
The constraint is expressed as
\begin{eqnarray}
Q_+^2=2aM_-(\xi -1),
\end{eqnarray}
where $\xi$ is an asymmetry parameter defined by $\xi \equiv M_+/M_-$.
The reflection-asymmetric thin-shell wormhole with a double shadow must have the throat in domains $r_{\mathrm{EH}-}<a<r_{\mathrm{PS}-}$ and $r_{\mathrm{EH}+}<a<r_{\mathrm{PS}+}$.
Permitted parameters $(\xi, a/M_-)$ for the reflectional-asymmetry thin-shell wormhole with the double shadow are shown in Fig.~1.
\begin{figure}[htbp]
\begin{center}
\includegraphics[width=85mm]{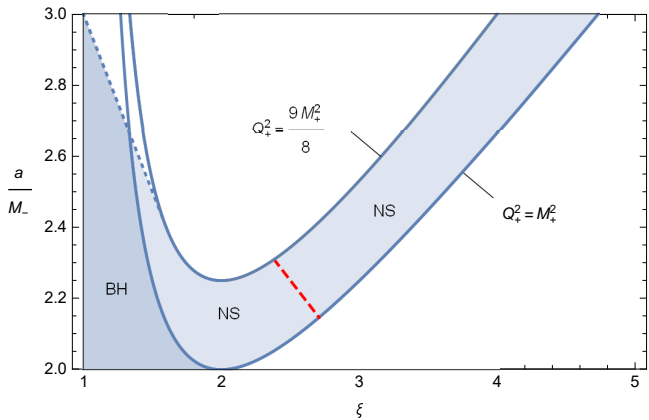}
\end{center}
    \caption{Permitted parameters $(\xi, a/M_-)$ of the asymmetric wormhole with the double shadow are in shaded zones. 
    $\mathcal{M}_+$ is the part of the Reissner-Nordstrom black hole (BH) spacetime in a deep blue shaded zone 
    while it is the part of the Reissner-Nordstrom naked singularity (NS) spacetime in light blue shaded zones.
    $\mathcal{M}_-$ is the subset of the Schwarzschild black hole spacetime.
    A blue dotted line denotes $a=r_{\mathrm{PS}+}$. 
    A red dashed line is explained in Sec.~IV.}
\end{figure}

\section{Stability of thin-shell wormhole}
We consider the linearization stability of a static wormhole with a thin shell at $a=a_0$ under the constraint (\ref{eq:constraint}) and $Q_-=0$.
The surface energy density $\sigma_0$ and pressure $p_0$ of the thin shell are given by 
\begin{equation}
\sigma_0= - \frac{\sqrt{f_{0}}}{2\pi a_0}
\end{equation}
and 
\begin{eqnarray}
p_0=\frac{1}{8\pi \sqrt{f_{0}}} \left( \frac{2f_{0}}{a_0} +\bar{f}^{\prime}_{0} \right),
\end{eqnarray}
respectively.
Here and hereafter a function with subscript~$0$ means the function at $a=a_0$.
Since $V_0=V^{\prime}_0=0$ is satisfied, 
the effective potential can be expanded around $a=a_0$ as
\begin{equation}
V(a)=\frac{V_0''}{2}(a-a_0)^2 +O\left( \left( a-a_0 \right)^3 \right), 
\end{equation}
where $V_0''$ is given by
\begin{equation}\label{eq:ddV_0}
V_0''
=A_0- B_0 \left( 1+2\beta^2_0 \right),
\end{equation}
where $A_0$ and $B_0$ are defined by
\begin{equation}
A_0 \equiv  \bar{f}^{\prime \prime}_0 -\frac{\Delta^{\prime 2}_0}{2f_0} -\frac{\bar{f}^{\prime 2}_0}{2f_0} 
\end{equation}
and 
\begin{equation}
B_0 \equiv \frac{2f_0}{a_0^2}-\frac{\bar{f}^{\prime}_0}{a_0},
\end{equation}
respectively.
The thin shell is stable (unstable) when $V_0''>0$ $(V_0''<0)$. 
Thus, the thin shell is stable when
\begin{equation}
\beta^2_0<\frac{1}{2}\left( \frac{A_0}{B_0}-1 \right)
\end{equation}
and $B_0>0$ holds
or 
\begin{equation}
\beta^2_0>\frac{1}{2}\left( \frac{A_0}{B_0}-1 \right)
\end{equation}
and $B_0<0$ holds.
Figure 2 shows the parameter zone of $(\xi, a_0/M_-)$ for $B_0>0$, the one for $B_0<0$, and their boundary $B_0=0$.
\begin{figure}[htbp]
\begin{center}
\includegraphics[width=85mm]{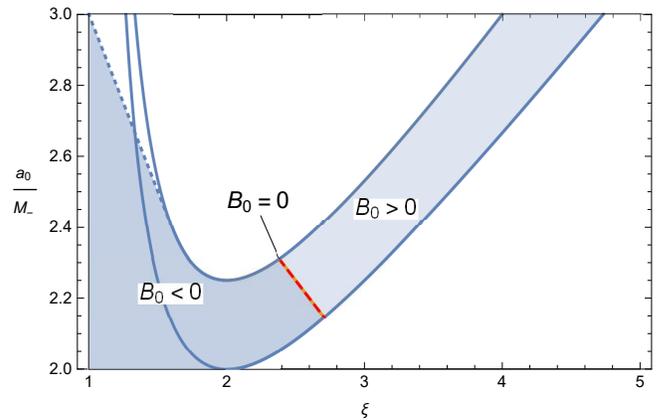}
\end{center}
    \caption{$B_0<0$ holds in a deep blue shaded zone and $B_0>0$ holds in a light blue shaded zone.
    The red dashed line $a_0/M_-=(7-\xi)/2$ for $(32+6\sqrt{2})/17 \leq \xi \leq 2+ \sqrt{2}/2$ denotes the boundary $B_0=0$.}
\end{figure}
The boundary $B_0=0$ is given by 
\begin{equation}
a_0/M_-=(7-\xi)/2 \quad  \mathrm{for} \quad \frac{32+6\sqrt{2}}{17} \leq \xi \leq 2+ \frac{\sqrt{2}}{2}.
\end{equation}
On the boundary, the thin shell is unstable for any $\beta_0^2$.
The parameters $(a_0/M_-, \beta^2_0)$ for the stable thin shell are shown in Fig.~3.
\begin{figure*}[htbp]
\begin{center}
\includegraphics[width=85mm]{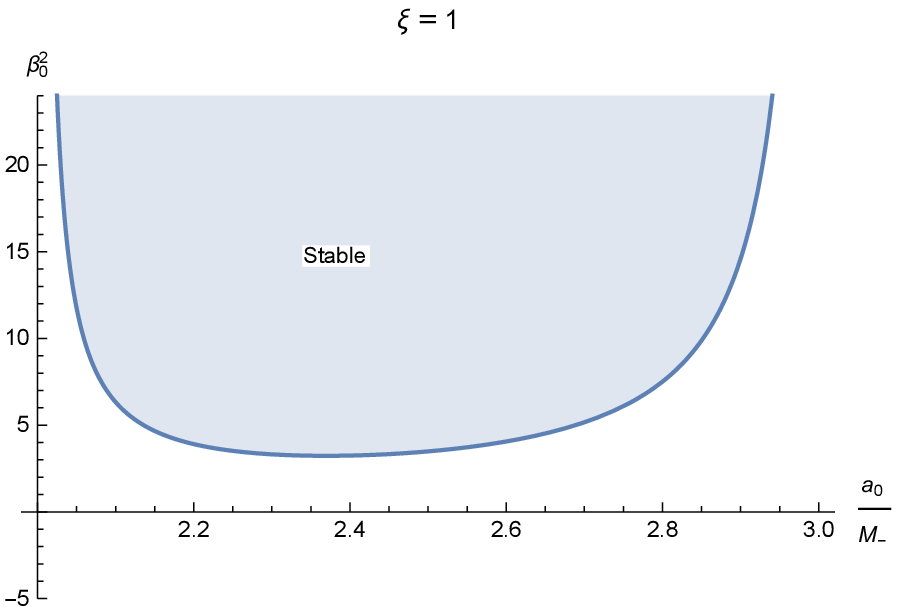} 
\includegraphics[width=85mm]{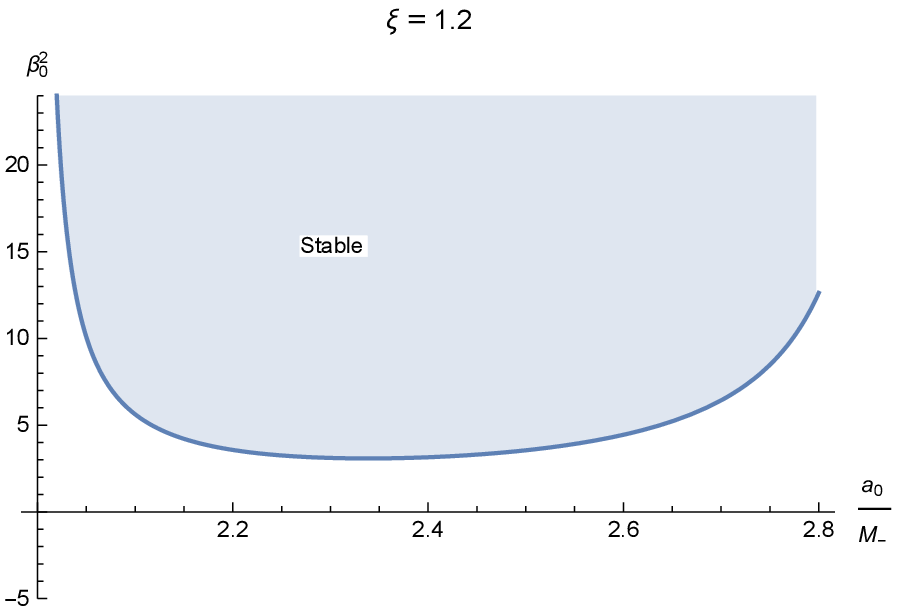} 
\includegraphics[width=85mm]{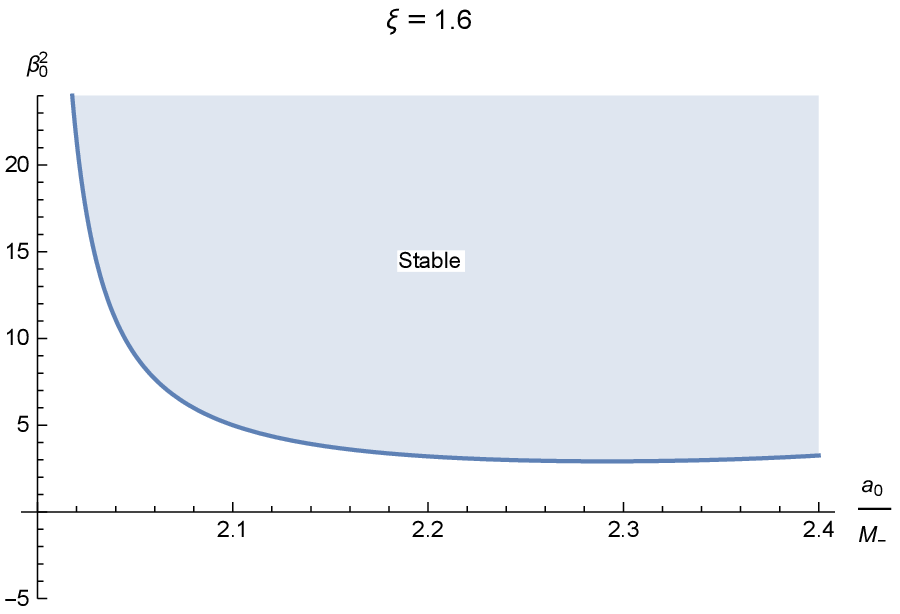}
\includegraphics[width=85mm]{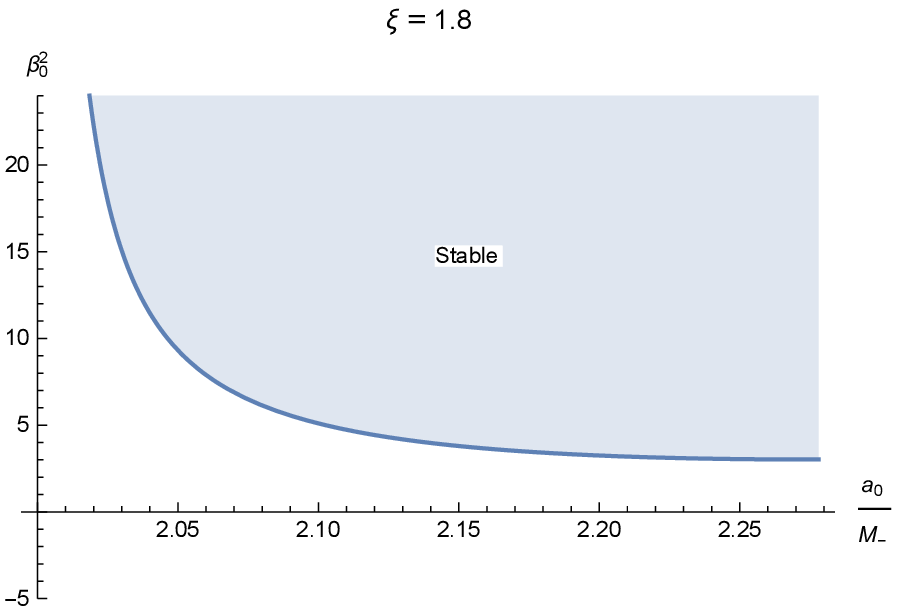}
\includegraphics[width=85mm]{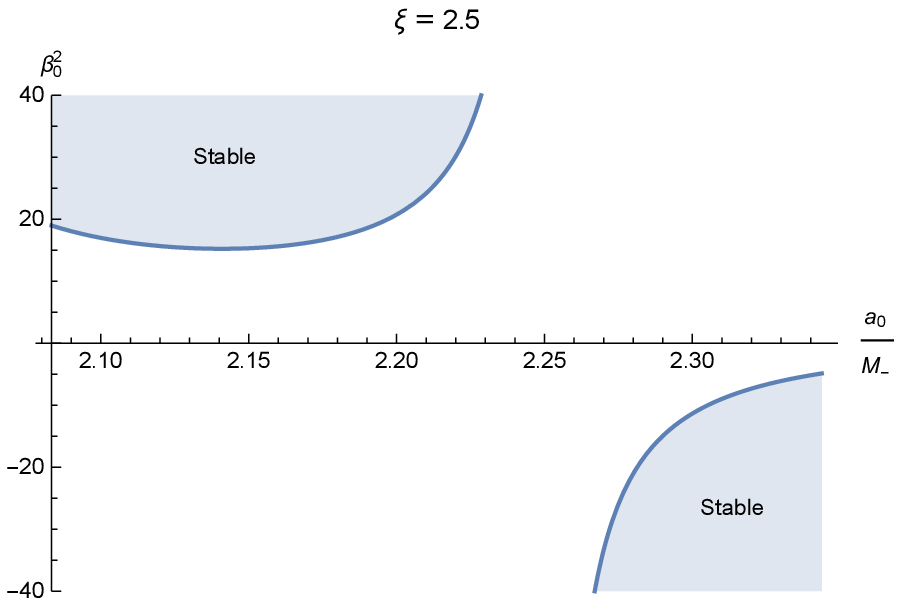}
\includegraphics[width=85mm]{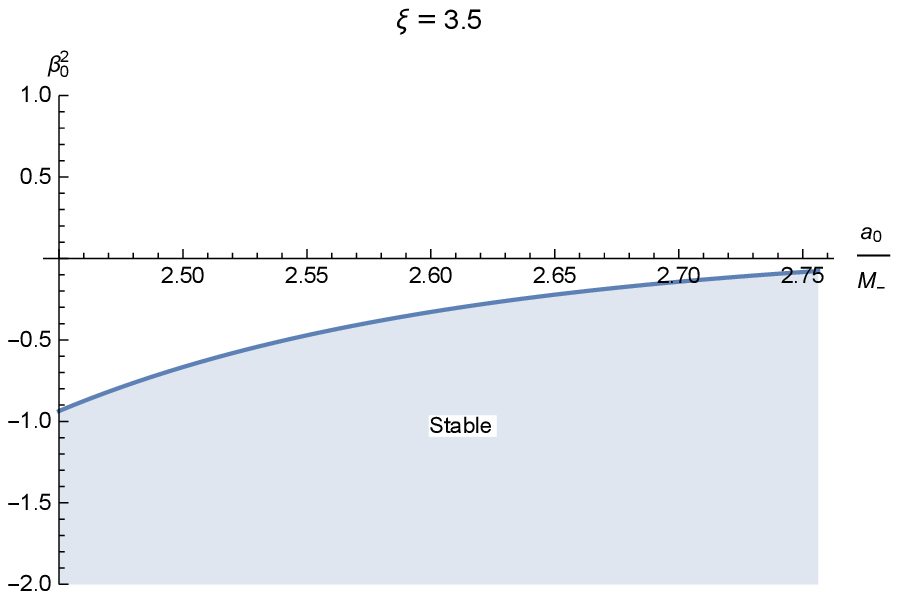}
\end{center}
    \caption{The static thin shell is stable if it is in blue shaded zones of $(a_0/M_-, \beta^2_0)$. 
    Left top, right top, left middle, right middle, left bottom, and right bottom panels show the cases of 
    $\xi=1.0$ for $2<a_0/M_-<3$, $\xi=1.2$ for $2<a_0/M_- < 2.8$, $\xi=1.6$ for $2 < a_0/M_- < 2.4$,
    $\xi=1.8$ for $2< a_0/M_- < 729/320$, $\xi=2.5$ for $25/12 < a_0/M_- < 75/32$, and
    $\xi=3.5$ for $49/20 < a_0/M_- < 441/160$, respectively.}
\end{figure*}

\section{Conclusion and Discussion}
We have investigated the linearization stability of the reflection-asymmetric thin-shell wormhole 
which is composed of the parts of the Schwarzschild and Reissner-Nordstrom manifolds with the double shadow.
We have imposed the constraint (\ref{eq:constraint}) as well as Ref.~\cite{Wielgus:2020uqz}.
The linearization stability given by Fig.~3 is characterized by 
the boundary $B_0=0$ or $a_0/M_-=(7-\xi)/2$ for $(32+6\sqrt{2})/17 \leq \xi \leq 2+ \sqrt{2}/2$ shown as the red line in Fig.~2.
In the boundary case $B_0=0$, the throat with $a_0/M_-=2.25$ is unstable for any $\beta^2_0$ as shown the left bottom panel with $\xi=2.5$ in Fig.~3.
We also note the throat at $a_0/M_-=r_{\mathrm{EH}-}/M_-=2$ for $1 \leq \xi \leq 2$ is unstable for any $\beta^2_0$ as shown the panels in Fig.~3.

We comment on the Schwarzschild thin-shell wormhole with reflection symmetry or $\xi=1.0$.
In Fig.~2, $B_0$ vanishes at a point $(\xi, a_0/M_-)=(1.0, 3.0)$.  
Thus, the throat of the Schwarzschild reflection-symmetric wormhole with $\xi=1.0$ at $a_0/M_-=3$ is unstable for any $\beta_0^2$ as shown Fig.~3.

We also give comments on wormholes without a thin shell while we have concentrated on the thin-shell wormhole on this paper.
The earliest traversable wormhole filled with a phantom scalar field~\cite{Martinez:2020hjm} was investigated 
by Ellis~\cite{Ellis:1973yv} and Bronnikov~\cite{Bronnikov:1973fh}. 
The Ellis-Bronnikov wormhole is a reflection-asymmetric wormhole when it has 
a positive Arnowitt-Deser-Misner (ADM) mass in a side and a negative ADM mass in another side
and it is a reflection-symmetric wormhole when it has vanishing ADM masses in the both sides.
See Refs.~\cite{Ellis:1973yv,Nandi:2016uzg} for gravitational lensing by the Ellis-Bronnikov wormhole with the positive ADM mass, 
see Refs.~\cite{Cramer:1994qj,Bronnikov:2018nub} for the negative ADM mass,
and see Ref.~\cite{Tsukamoto:2016qro} and references therein for the vanishing ADM mass.
The instability of the Ellis-Bronnikov wormhole was reported~\cite{Shinkai_Hayward_2002,Gonzalez_Guzman_Sarbach_2008_I,Gonzalez_Guzman_Sarbach_2008_II}.
Reliable stable wormholes without thin shells have not reported yet in general relativity
but we may construct stable wormhole solution by choosing exotic matters other than the phantom scalar field~\cite{Bronnikov:2013coa}. 
Recently, Shaikh~\textit{et al.} have suggested gravitational lensing by reflection-symmetric wormholes with multiple photon spheres 
and the shadow images of the multiple photon spheres~\cite{Shaikh:2018oul} and 
Bronnikov and Baleevskikh have considered that all asymptotically-flat, static, spherically symmetric wormholes with reflection symmetry have 
a photon sphere on a throat~\cite{Bronnikov:2018nub}. 
However, we notice that they overlook the possibility that circular photon orbits on the throat can form an antiphoton sphere. 
The gravitational lensing by wormholes with the antiphoton sphere on the throat has been suggested by Shaikh~\textit{et al.} in Ref.~\cite{Shaikh:2019jfr}.
Tsukamoto has shown that
the circular photon orbits on the throats of a Damour-Solodukhin wormhole~\cite{Damour:2007ap} and a Simpson-Visser wormhole~\cite{Simpson:2018tsi} 
can be the antiphoton spheres~\cite{Tsukamoto:2020uay,Tsukamoto:2020bjm}. 
In Ref.~\cite{Bronnikov:2018nub}, Bronnikov and Baleevskikh 
have discussed the deflection angle of a light ray scattered by reflection-asymmetric wormholes with the photon sphere which is slightly off the throat. 
Gravitational lensing by reflection-asymmetric wormholes with the antiphoton sphere which is slightly off the throat is left as a future work.

\textit{Note added.} Recentely, related papers have appeared on arXiv.
Guerrero \textit{et al.} have constructed reflection-asymmetric wormholes supported by a positive energy thin shell 
with a double shadow in Palatini $f(R)$ gravity~\cite{Guerrero:2021pxt}.
Peng \textit{et al.} have studied the observational appearance of an accretion disk around a reflection-asymmetric thin-shell wormhole~\cite{Peng:2021osd}.

%

\end{document}